\documentclass[a4paper,12pt]{JHEP3}
\usepackage{amssymb}
\usepackage{mathrsfs}
\usepackage{cite}
\usepackage{texdraw}
\usepackage[english]{babel}
\usepackage{epsfig}

\csname @addtoreset\endcsname{equation}{section}

\newcommand{\HH}{\mathcal{H}}

\newcommand{\bra}[1]{< #1|}
\newcommand{\ket}[1]{| #1>}

\newcommand{\bea}{\begin{eqnarray}}
\newcommand{\eea}{\end{eqnarray}}

\title{Rational foundation of GR in terms of statistical mechanic
in the AdS/CFT framework}
\author{Pedro J. Silva \\
Dipartimento di Fisica dell'Universit\`a di Milano, and\\
INFN, Sezione di Milano\\
Via Celoria 16, I-20133 Milano, Italy \\
Institut de Fisica d'Altes Energies (IFAE),\\
E-08193 Bellaterra (Barcelona), Spain.\\
E-mail: \email{pedro.silva@mi.infn.it}}
\preprint{\\hep-th/0505xxx}

\abstract{In this article, we work out the microscopic statistical
foundation of the supergravity description of the simplest
1/2 BPS sector in the AdS(5)/CFT(4). Then, all the corresponding
supergravity observables are related to thermodynamical observables,
and General Relativity is understood as a mean-field theory. In particular,
and as an example, the Superstar is studied and its thermodynamical
properties clarified.}


\begin{document}

\section{Introduction}

The AdS/CFT correspondence is a powerful framework to study the
microscopic structure of space-time \cite{Maldacena:1997re}. In
this setting the gravitational field theory can be regarded as a
{\it mean-field theory}, resulting from a coarse graining of the
microscopic degrees of freedom. In this form, the role of General
Relativity is more alike to an {\it equation of state}, a derived
effective theory \footnote{Here, we follow Wilson ideas to define
an effective theory, that is obtained by integrating out
ultraviolet degrees of freedom up to a given scale, that sets the
range of validity of the resulting effective theory.}, that is
obtained from thermodynamical arguments. Note that from this point
of view, its direct quantization is as useful as the quantization
of an effective theory, and therefore has little to tell about the
ultraviolet degrees of freedom that have been integrated out. In
fact, it was notice in \cite{Jacobson:1995ab} that from
generalized Black Hole entropy laws, Einstein equations can be
deduce. Of course, this last point does not prove our point of
view, but it certainly reinforce it and show its consistency.

In this setting, we have to face the fact that observables in
general relativity are all intrinsically thermodynamical, and that
our main quest should be to find a statistical mechanics arguments
to give a {\it rationale foundation} to General Relativity.

In particular, Black Holes (BH) should be seen in the same light.
Here General Relativity shows more explicitly its thermodynamical
aspects (e.g. the BH thermodynamical laws) and its limits, like
the pathological singularities at its core or the possible lost of
unitarity. Following the above line of reasoning, we have to
revised our concepts of BH physics, to find out up to what extent
we can push the classical picture (like for example the concept of
event horizon), into quantum process. At this point, we have to
make reference to a particular incarnation of this framework, in
the work of Mathur et. al. ( see\cite{Lunin:2002bj} and reference
there in), where the actual definition of a BH is replaced by a
sort of coarse grained geometry, where the meaningful concept
seems to rely on the emergence of semiclassical microstates
geometries. Here, the event horizon and basically any classical
observable are just a blur picture resulting from a measurement of
a system in a mixed state in the underlying ultraviolet partons.
Nevertheless, we should also recalled that this particular
incarnation is not fully understood and checked for realistic BH,
and more work needs to be done to bring it into more firm grounds.
In any case, it is clear the BH physics begs for a rational
foundation in terms of statistical arguments of the ultraviolet
degrees of freedom.

A particulary useful sector in the AdS/CFT duality is the one that
deals with the simples 1/2 supersymmetric configurations. In the
CFT side it is describe by a Matrix quantum mechanics that can be
recast in terms of many non-interacting fermions in a quantum Hall
effect, coffined to stay in the lowest Landau level (see
\cite{Corley:2001zk,yo}). The Gravity side is describe by the LLM
construction \cite{Lin:2004nb}, where only the metric and the
Ramond-Ramond 5-form are switched on. The beauty of this sector is
that the classical limit of the CFT theory (i.e. the form of the
corresponding droplets in phase space) sources the space-time
solutions, in a one to one correspondence. Unfortunately this
sector does not contain any BH, but a poor relative called
superstar \cite{Myers:2001aq} (SS), with no horizon and a naked
singularity at its center. Nevertheless, the SS develops a horizon
as soon as it is perturbed by non bps states, and it has been
argue in \cite{1} that it actually grows a small horizon after
stringy corrections are taken into account. In any case, this
solution is the responds of space-time to a distribution of
D3-branes (in this context named Giant gravitons), that has an
associated degeneracy of states that scales exponentially with its
energy and therefore gives a non vanishing entropy.

{\it Here we study a model, based on statistical mechanics of the
dual CFT system of the LLM sector, to give a rational foundation
of its thermodynamical properties or as it was formally known, its
GR description}. For definiteness, we study the statistical
properties of all the states in the CFT Hilbert space, of fixed
Energy, that source the SS as an example. Therefore we work in the
Microcanonical ensemble. To make a connection to the classical
phase space we use the phase space formulation of quantum
mechanics of Weyl \cite{weyl}and Wigner \cite{wigner} and the
associated inverse operators of Groenewold \cite{groenewold}.

The result of the study are a clear picture of the emergence of
the mean-field description, underlying the thermodynamical
properties of the SS and in general extendable to the whole 1/2
BPS sector.

We are aware of two works where the SS has been studied using the
fermion picture, \cite{1,2}. In the first work, the entropy of the
dual system to the SS is calculated in a particular regime, to
conclude that is consistent with the appearance of an stretch
horizon (following Mathur ideas), defined by the microstate with
lower energy. In the second article, it is shown the the naive
Microcanonical ensemble does not reproduced the correct
semiclassical density of states and that the definition of the
stretch horizon is subtle. Here, we find a solution to the
ensemble problem, giving the form of the distribution, leaving
open the definition of the will-be stretched horizon.

Also, the idea that General relativity is an effective theory, is
really a definition within string theory. Here, we are giving a
explicit scenario, with examples, were the quantum nature of the
microscopic degrees of freedom are studied and the thermodynamical
nature of General Relativity observed with clarity. There are
other two works with similar intentions \cite{4,3}. The first one
using a formalism that is different, recovers the semiclassical
vacuum density and also propose a method to study less symmetric
sectors in the AdS/CFT (i.e. with less supersymmetry). The second
is a titanic effort to understand the foundations of GR, and
certainly study in detail the SS. Nevertheless, their approach is
again different to our approach. In \cite{3} a connection between
the semiclassical form of the fermionic droplets and the form of
Young diagrams representing the ensemble of fermions is
established (basically both are pictorial representations of
energy dependent ensembles of states).

The articles \cite{2,4,3} appeared while we where at the final
stage of this work. There is overlap and we discuss similar
matter, although our focus is different. We believe that our point
of view is valuable, and adds for a more complete understanding of
this matters.


\section{Quantum Mechanics, Phase Space and Classical limit}
\label{a}

The basic idea we are using, is that the semiclassical limit of
the different states of $N$ non-interacting fermions, are the
boundary conditions the uniquely defines the dual gravity
configuration. To be more precise, in the semiclassical limit
fermions have a minimal area of order $\hbar$ in phase space. Due
to Pauli exclusion principle, they can not be at the same point,
therefore spread forming droplets. This droplets are in turn the
boundary data that specifies uniquely the dual supergravity
configuration. It is important to make notice that many different
quantum states may form the same classical droplet, and in fact,
in the correspondence it is used that measurements of order
$\hbar$ are not possible to be resolved. This is the essence for
the coarse graining picture.

In order to understand the duality, we need to get a grip of the
classical limit of the fermion system in term of phase space
variables incorporating the notion of ensembles of states, since
the supergravity measurements will in general, only define a
subspace of the total Hilbert space. To this end, we use Weyl and
Wigner formalism, where in short, quantum mechanics is rewritten
in term of a phase space operator density.

 \vspace{.5cm}\noindent \textsc{\bf One particle case}
 \vspace{.5cm}

For simplicity, we start with a single particle Hilbert space,
that gives a two-dimensional phase space. A particulary useful
operator to investigate on thermodynamical properties is the
density operator $\hat\rho$. For a mixed state in a n-dimensional
subspace of the Hilbert space, we have that
\bea
\hat \rho =\sum_n
P_n\ket{\psi_n}\bra{\psi_n}\quad\quad \sum_n P_n =1,,
\eea
where the
expectation value of an operator $\hat A$ is given by $<\hat A>=
Tr(\hat A \hat\rho)$.

Then, given an operator $\hat A$ the Weyl transformation
associates to it, a function $A(q,p)$ as follows
\bea
\label{weyl}
A(q,p)=\int dy \bra{q+\hbox{${y\over2}$}}\hat
A\ket{q-\hbox{${y\over2}$}}e^{-ipy/\hbar}\,,
\eea
where the $\ket{q}$ is the usual position basis. Then,
the Wigner density is just the Weyl transformation of $\hat\rho$,
\bea
W(q,p)={1\over 2\pi\hbar}\int dy \bra{q+\hbox{${y\over2}$}}\hat
\rho\ket{q-\hbox{${y\over2}$}}e^{-ipy/\hbar}\,,
\eea

\DOUBLEFIGURE[t]{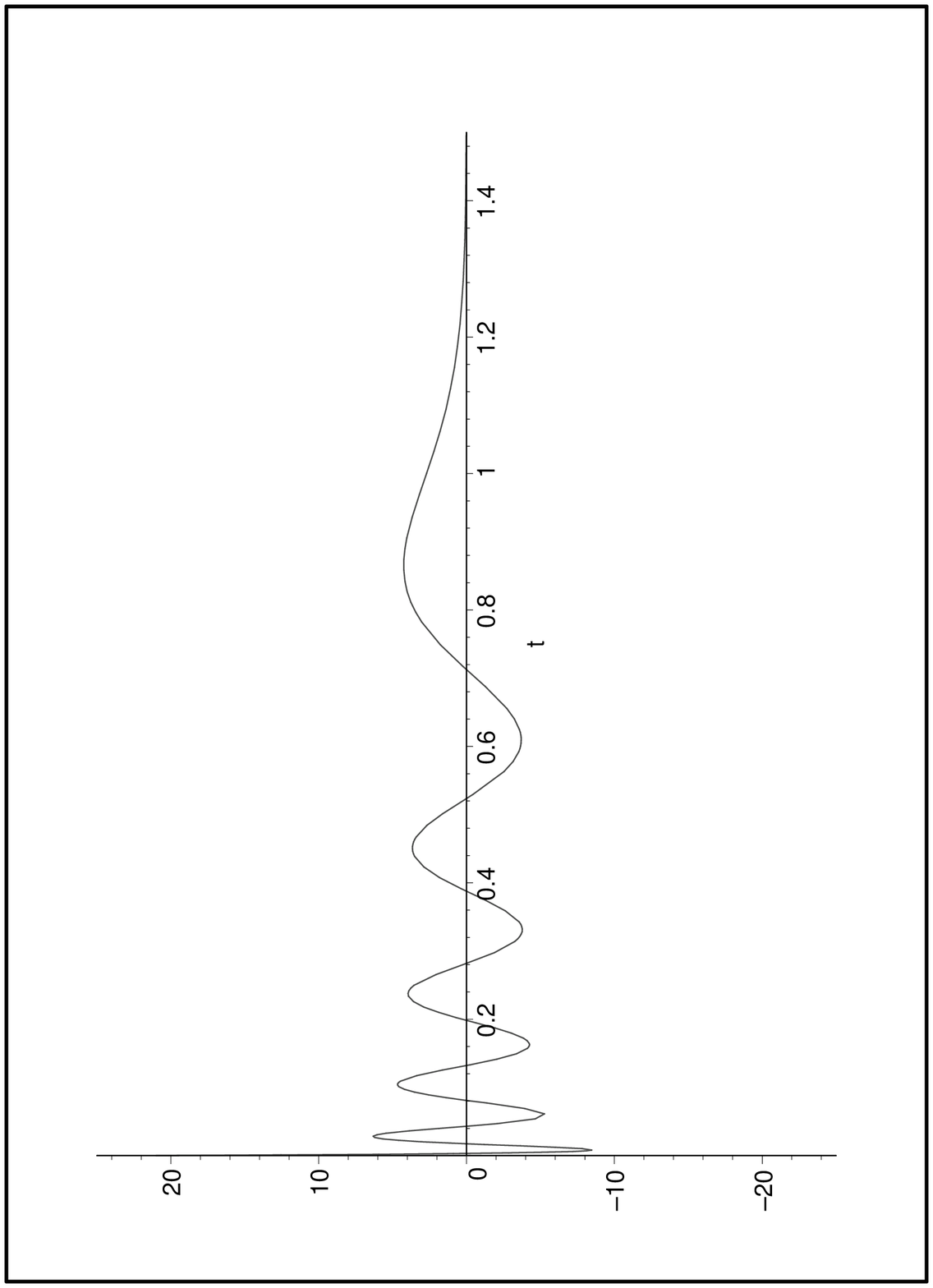,angle=-90,scale=.3}{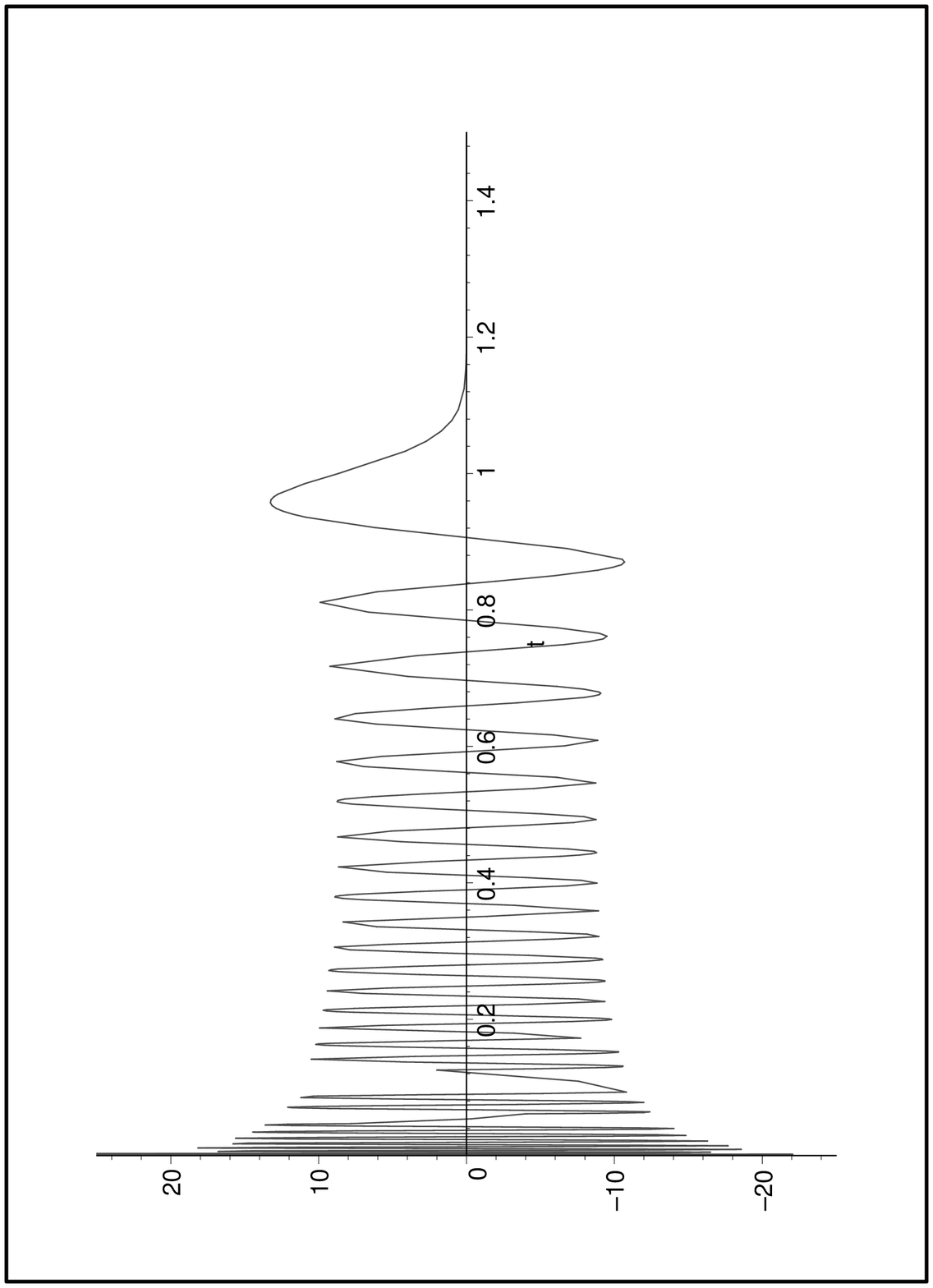,angle=-90,scale=.3}
{plot for n=10}{plot for n=40}

There are many interesting properties of $W$ that deserved
 attention but here we will just say that for Harmonic oscillators,
 its dynamics is identical to the dynamics of the classical
 Liouville density. On the other hand it is well known that this
 operator has negative eigenvalues, a feature not very pleasant for
 a candidate to a classical density. Nevertheless, we are interested in the
 semiclassical limit of a harmonic oscillator, and here, $W$
 behaves much more nicely. Consider the harmonic oscillator,
 with Hamiltonian $\bar H=p ^2/2m + (mw^2/2)q ^2$, with
 energy eigenstates $\ket{n}$ and energy levels $E_n=\hbar w(n+1/2)$.
 then $W$ for one of these pure states is given by
 \bea
W(n;q,p)=\left[(-1)^n\over \pi\hbar\right]e^{[-2H(q,p)/\hbar
w]}L_{(n)}[4H(q,p)/\hbar w]\,,
 \eea
where $L_{(n)}$ is the Laguerre polynomial of order $n$. To study
the classical behavior is useful to define the dimensionless
quantity $\tau= {2H \over \hbar w} = \left(\beta^2 q^2+
\hbox{${1\over\hbar^2\beta^2}$}p^2\right)$ with $\beta^2 = {mw\over \hbar}$.
Then we have that
\bea
\lim_{\hbar\rightarrow 0}W(n;p,q)= (1/\pi)\delta(t)
\eea
This is expected since in this limit $E_n\rightarrow 0$. Nevertheless we
are interested in a different classical limit where the energy of
our states $E$, is kept constant\footnote{In the AdS/CFT sector we
are interesting to study, this limit corresponds to maintain
constant the size of the droplets while the minimal are of $\hbar$
size is send to zero.}. In this case, we have to send
$n\rightarrow \infty$ as $\hbar\rightarrow 0$. This is a more
subtle limit and care should be taken to understand what is going
on. For large $n$ and small $\hbar$ we get a rapidly oscillatory
behavior, and a final lump at the value of the fixed energy
$t_0=E_0/w$ (see fig. 1 and fig.2 where we plot the Wigner density for
different values of $n$).

To preform a measurement of an
observable $A$, we have to integrate $W\times A$ on phase space.
Basically, the oscillatory part of $W$ cancels out, and only the
last lump gives a meaningful contribution. Therefore we can write
to good approximation that
\bea
\lim_{E fixed}W(n;p,q)=
(1/\pi)\delta(t-t_0)\,.
\eea

Summarizing, for a single particle in a harmonic potential, the
Wigner density reproduce the desiderate classical limit when
$\hbar$ goes to zero and the Energy is fixed constant. In
particular we have only consider the energy, and therefore, our
classical density defines a ring on phase space. To talk about
more localized classical limits in phase space, coherent states
should be used\footnote{see \cite{yo} where some of those coherent
states where constructed to describe giant graviton.}. The
presented developments are enough for the scope of this work.

\vspace{.5cm}\noindent \textsc{\bf Many particles case}
\vspace{.5cm}

Consider now, the case of $N$ fermions. This time, the Hilbert
space is the tensor product of $N$ single particle Hilbert spaces,
$\HH_{total}=\HH_1\otimes \HH_2\ldots\otimes \HH_N$. Due to the
Dirac statistics, any physical state has to be anti-symmetric on
each single particle Hilbert space. Using the creation and
annihilation operators $a^{\dag}_i, a_i$ acting on $\HH_i$, a
basis for the physical states, can be label by a N-dimensional
vector $\vec{n}$ where its coordinates $n_i$ corresponds to the
number of times the creation operator $a^\dag _i$ acts on the
vacuum. In this notation, anti-symmetrization is understood, where
$n_i>n_j$ if $i<j$ and when written explicitly the Slater
determinant of the following form is obtained
\bea
  \ket{\vec{n}}= {\det\pmatrix{
      &({a^\dagger_1})^{n_1} & ({a^\dagger_1})^{n_2} & \cdots & ({a^\dagger_1})^{n_N} \cr
      & ({a^\dagger_2})^{n_1} & ({a^\dagger_2})^{n_2} & \cdots & ({a^\dagger_2})^{n_N} \cr
      & \vdots &; \vdots &; \ddots &; \vdots \cr
      & ({a^\dagger_N})^{n_1} & ({a^\dagger_N})^{n_2}& \cdots & ({a^\dagger_N})^{n_N} \cr }
  }
  \ket{0}. \\
\nonumber
\eea
The above
notation also has a closed connection to $U(N)$ Young diagrams
where each $n_i$ is associated to the number of boxes of on the
row $"i"$, in a given diagram. Note that each of these states has
a definite energy $E_n=\hbar w\sum_{i=1}^{N}(n_i+1/2)$, and the energy
of the fermionic vacuum or Fermi sea is $\hbar w{N^2\over 2}$.

Let us write the form of the Wigner density for one of the above
states $\ket{\vec n}$, following our previous definitions we write
\bea
W(\vec n, \vec q, \vec p)={1\over 2\pi\hbar}\int (dy)^N
\bra{\vec q+\hbox{${\vec y \over 2}$}}\vec n><\vec n\ket{\vec
q-\hbox{${\vec y \over2}$}}e^{-i\vec p \vec y /\hbar}\,,
\eea
where the integration is taken over an $N$-dimensional space, and
therefore $W$ is a function of the vectors $(\vec q, \vec p)$.

To have an idea of the form of $W$, let´s consider the Fermi vacuum
state for $N=2$. In this case the vacuum is given by
\[
\ket{1,0}={1\over \sqrt 2}(a^{\dag}_1-a^{\dag}_2)\ket{0} \quad,\quad
\bra{\vec q+\hbox{${\vec y \over 2}$}}\vec n>={\beta\over 2\sqrt \pi}\,
e^{-{\beta^2\over 2}\,\vec{q}^2}[H_1(\beta q_1)-H_2(\beta q_2)]
\]
and hence, after some algebra we get
\bea
W(v_{2};\vec q,\vec
p)=-\left(2\over\pi\hbar\right)e^{-\left(\beta^2 \vec{q}^2+
\hbox{${1\over\hbar^2\beta^2}$}\vec{p}^2\right)}
\left\{1-\left[\beta^2 (\Delta q)^2+{1\over\hbar^2\beta^2}(\Delta
p)^2\right]\right\},
\eea
where the $v_{2}$ stand for the
two-dimensional vacuum and $\Delta a= a_1-a_2$. At this point, to
recover the density function, depending only on one pair of
canonical coordinates $(q,p)$, we just have to integrate over the
other canonical pairs, since what we want to define, is the
probability to measure a fermion, doesn't matter which one, in a
given position in phase space. After integration, with proper care
on the anti-symmetrization properties, we get
\bea
W(v_{2};q,p)={1\over \pi\hbar}\left[
e^{-\tau}-e^{-\tau}(1-\,2\tau)\right],
\eea
where $\tau$ is
defined as in the single particle case. The above calculation when
generalized to $N$ fermions gives
\bea
\label{wv}
W(v_{N};q,p)=
{1\over 2\pi\hbar}\sum_{n=0}^{N-1}(-1)^n2e^{-\tau}L_n(2\tau)
\eea

where the $v_{2}$ stand for the N-dimensional vacuum. This same
expression can be found by a complete different approach, where the
quantization program is based on a non-commutative star product $*$
(see for a review with application to this picture
\cite{Dhar:2005qh}). In this framework $W$ has to satisfies the
equation
\[
W*W(p,q)=W(p,q)
\]
where
\[W*W(p,q)=e^{i\hbar/2(\partial_{q1}\partial_{p2}-\partial_{q2}\partial_{p1})}
W(q_1,p_1)\,W(q_2,p_2)|_{(q_1=q_2=q;p_1=p_2=p)}.\]

Now, in the classical limit where the higher energy level is kept
constant, this Wigner density produces a step function, that
translates into a circular droplet in phase space. A insightful
way to understand this is from the single particle picture, as
follows: The classical limit is taken, in such a way that the
higher energy level remains constant i.e. the $\hbar\sim N^{-1}$
while $N\rightarrow \infty$. The last term of the sum in equation
(\ref{wv}), behaves more and more like a delta function center at
the corresponding $\tau_0$, but the other terms also tend to delta
functions, only that they are centered at lower $\tau_i$, forming
a chain of delta functions, one after the other. In the large $N$
limit, they become indistinguishable from a step function, since
we can not resolve distances of order $1/N$ (see fig. 3 and fig .4).

\DOUBLEFIGURE{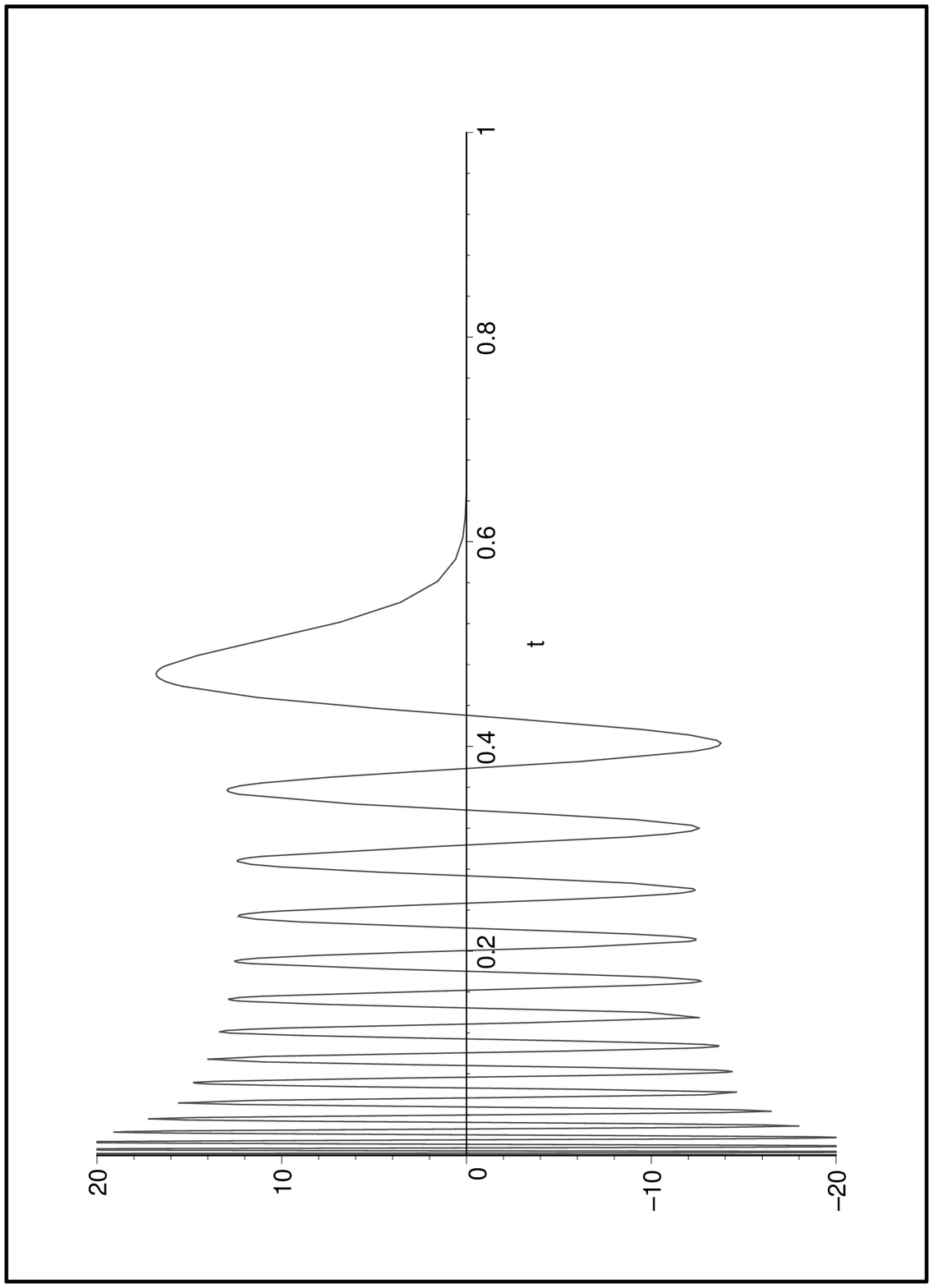,angle=-90,scale=.3}{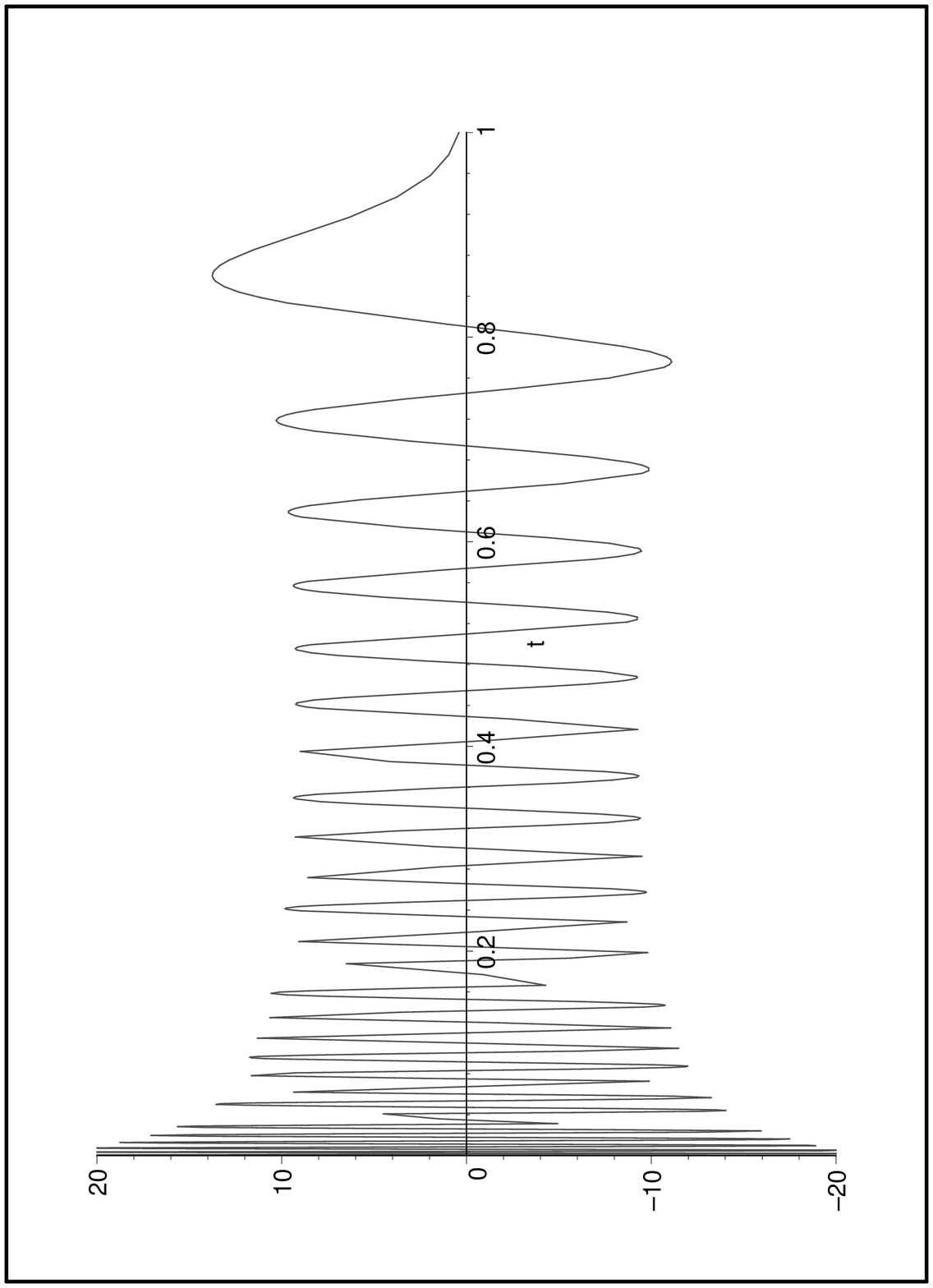,angle=-90,scale=.3}{The delta
distribution is center at $\tau=30/60$ }{The delta
distribution is center at $\tau=54/60$}

Therefore we see the emergence of the vacuum configuration out of
the quantum theory in a very explicit way. Each term of the sum
should be identified with a given fermion, since in the classical
limit these terms collapse into $N$ delta functions, forming a circular
droplet out of infinitesimal concentric rings.

At this point, we are ready to write the form of the Wigner
function, for general pure states of the form $\ket{\vec n}$. The
result is clearly a sum of $N$ terms of the form
$e^{-\tau}L_{n_i}(2\tau)$ where each one has the same coefficient,
i.e.
\bea
\label{wg}
W(\vec n ;q,p)= {1\over
2\pi\hbar}\sum_{n_i}(-1)^{n_i}2e^{-\tau}L_{n_i}(2\tau)\,.
\eea
This picture is in complete agreement with the slater determinant
picture studied in \cite{Corley:2001zk,yo}. For example, the
Wigner density of a giant graviton with energy $m$, growing in AdS
(corresponding to a schur polynomial in the symmetric
representation of $S_N$ of degree $m$ with vector
$\vec{n}_m=(m+N-1,N-2,\ldots,0)$) has the expected form
\bea
\label{wgga}
W(\vec{n}_m;q,p)={1\over
2\pi\hbar}\sum_{n=0}^{N-2}(-1)^n2e^{-\tau}L_n(2\tau) + {1\over
2\pi\hbar}(-1)^{N-1}2e^{-\tau}L_{m+N-1}(2\tau)
\eea which also
agrees with the picture of concentric rings, exited above the
fermi sea level.

Another very interesting feature that we point out,
is that the $N$-Fermion Wigner function can be
understood as a {\it mixed state of a single particle}, where the
coefficient of the sum in equation (\ref{wv}), correspond to the
probability of occurrence of each single pure state with a funny
normalization.
\bea
\label{wm}
W(v_{N};q,p)=\sum_{n=0}^{\infty}P_n{1\over\hbar}(-1)^ne^{-\tau}L_n(2\tau)\quad,
\quad \sum_{n=0}^{\infty}P_n=N\,.
\eea

Summarizing, as in the single particle case, we have been able to
reproduce the classical limit of the Wigner density for the
$N$-fermion system. In this case, The final form of the density is
obtain by integrating out the extra $(N-1)$ canonical pair of
coordinates $(q_i,p_i)$. The resulting function is the
superposition of $N$ single particles in different eigenstates of
the energy, defined by either its Young diagrams, or the vector
$\vec n$. We have reproduced in the classical limit, the vacuum
configuration corresponding to the round droplet, and also all
the droplets form by the superposition of concentric rings.
Also, the above pure states can be reinterpreted as a Wigner density
of a mixed state for a single particle. {\it This interpretation justifies
and translates into the superposition properties characteristic
of the GR regime, where the solutions are found by superposition of
different droplets configurations}.

\section{Statistical analysis and GR}

One of the more interesting outcomes of the previous analysis was
that the Wigner function for $N$ Fermions, can be reinterpreted as
a Wigner function of a single particle in a mixed state. As long a
we are in the quantum regime this is no more than a nice
observation. Nevertheless, as soon as we take the classical limit,
and the number of fermions grows, nearby fermions forms
indistinguishable drops, and many different initial configurations
result in the same final drop. Therefore, we lose the trace
of the true quantum state, to end up with a classical density that
only selects subspaces in the original Hilbert space. At this
point, due to our ignorance, we rather have an ensemble of states
or a mixed state. The fact that we interpret Wigner as a single
particle mixed state makes this transition smooth since or
statistical Wigner function of the classical system will be of the
same form, with a infinite sum or an integral if you prefer.

{\it Therefore our proposal is that the thermodynamical nature of
GR comes about due to the coarse graining of the classical limit,
that at the end, translates in our incapability to resolve the
specific quantum state that produces the classical observable.}

To test this conjecture, we study the nearest classical soliton in
this 1/2 BPS sector of supergravity to a BH, the superstar (SS).

\vspace{.5cm}\noindent \textsc{\bf Superstar as a mixed state}
\vspace{.5cm}

The SS solution has been much studied lately,
here we just need that it is a 1/2 BPS solution with energy
$E=N^2Q/2L=\Delta/L$, with a naked singularity an the center of
$AdS?5$, that extends all over $S^5$ and that has been interpreted as
the space-time responds of a particular distribution of giant
gravitons \cite{Myers:2001aq}.

From the point of view of the LLM construction, the above solution
corresponds to a circular drop of larger radius $R_Q$ than the
vacuum radius $R_0$, forcing the classical fermion density $\rho_c$,
to attain lower values than 1. To be more exact, $R_0^2=2\hbar N$,
and $R_Q^2=2\hbar N(1+Q)$,  in fact we have that
\[\rho_c={1\over 2\pi \hbar(1+Q)}\theta(r^2-R^2_Q)\,.\] In this case the
drop is said to be grey in contrast to black drops, where $\rho_c=1$.

We would like to see this gray density as the outcome of a Wigner
density in the Microcanonical ensemble\footnote{We chose the
Microcanonical ensemble for naturalness, due to our previous
discussions, but certainly other ensembles are valid, and indeed
has been used, see for example \cite{3}.}. Hence we write that
\bea
\hat \rho =
\sum_{i=1}^d P_i\ket{\vec {n}_\Delta}\bra{\vec {n}_\Delta}
\quad,\quad \sum_{i=1}^d P_i =1
\eea
where $\ket{\vec {n}_\Delta}$
are $N$-Fermion eigenstate of the Hamiltonian introduced before,
with total energy $\Delta$ and degeneracy $d$.

\vspace{.5cm}\noindent \textsc{\bf Equiprobable distribution}
\vspace{.5cm}

The particular form of $P_i$, depends on the character or the
statistical nature of the mixed state in the ensemble. A very
natural option will be the case where all $P_i$ are equal, i.e.
\[P_i=\left({1 \over d}\right)\,,\]
 nevertheless, it is not difficult to see that this "natural
 guess¨" can not be correct for al values of $Q$.
 To show this, consider the case when $Q<N$, and make notice that
 when Wigner density is computed, we get the following structure
 \bea
 W(\vec{n}_\Delta,q,p)={1\over 2\pi\hbar}\sum_{m=0}^{N-\Delta}
 (-1)^m2e^{-\tau}L_{2\tau}
  + {1\over 2\pi\hbar}\left({1 \over d}\right)\sum_d \delta W
  \eea
  where $\delta W$ is the contribution of the exited states over
 the Fermi sea of level $N-\Delta$ that is left unperturbed.
 In this case the corresponding classical density $\rho_c$
 would rather be a black droplet with an external ring of variable
 density (not equal 1)\footnote{In fact the observation that the
equiprobable distribution does not correspond to a grey droplet
has been discuss in \cite{2} from a different perspective.}.

What is going on, is that in the equiprobable case, there
are many possible ways to have a large number of giant gravitons
say $m$, with an energy of the same range, in comparison to the
case where we have a small number of very energetic giants or a
large number of giants with little energy. Basically, this tails of
the probability density spoil the result. What we need is a
distribution where the states that appear, have the same order of
energy and number of Giants. In terms of Young diagrams, we need
an ensemble of almost triangular diagrams\footnote{Of course the
equiprobable distribution could mimic the grey distribution in
some particular range for $Q$, where the above tales are
unimportant, but we would rather like to have a probability
density that works for al values of $Q$.}.

$$
  \btexdraw
  \drawdim cm
   \move(0 .5)\lvec(4 .5)
   \move(0 0)\lvec(4 0)
   \move(0 -.5)\lvec(3 -.5)
   \move(0 -1.5)\lvec(2.2 -1.5)
   \move(0 -2)\lvec(2.2 -2)
   \move(0 -2.5)\lvec(1.5 -2.5)
  \move(0 .5)\lvec(0 -2.5)
  \move(.5 -2.5)\lvec(.5  -1.5)
  \move(.5 -.5)\lvec(.5 .5)
  \move(1 -.5)\lvec(1 .5)
  \move(1   -.0)\lvec(1   .5)
  \move(1.5 -.0)\lvec(1.5 .5)
  \move(4 .0)\lvec(4 .5)
  \move(3 -.5)\lvec(3 0)
  \move(2.2 -2.0)\lvec(2.2 -1.5)
  \move(1.5 -2)\lvec(1.5 -2.5)
  \move(1.8 .25)\fcir f:0 r:.01
  \move(2.0 .25)\fcir f:0 r:.01
  \move(2.2 .25)\fcir f:0 r:.01
  \move(2.4 .25)\fcir f:0 r:.01
  \move(2.6 .25)\fcir f:0 r:.01
  \move(2.8 .25)\fcir f:0 r:.01
  \move(2.8 .25)\fcir f:0 r:.01
  \move(1.2 -.25)\fcir f:0 r:.01
  \move(1,4 -.25)\fcir f:0 r:.01
  \move(1.6 -.25)\fcir f:0 r:.01
  \move(1.8 -.25)\fcir f:0 r:.01
  \move(2 -.25)\fcir f:0 r:.01
  \move(.6 -1.75)\fcir f:0 r:.01
  \move(.8 -1.75)\fcir f:0 r:.01
  \move(1 -1.75)\fcir f:0 r:.01
  \move(.6 -2.25)\fcir f:0 r:.01
  \move(.8 -2.25)\fcir f:0 r:.01
  \move(.7 -.6)\fcir f:0 r:.01
  \move(.7 -.8)\fcir f:0 r:.01
  \move(.7 -1)\fcir f:0 r:.01
  \move(.7 -1.2)\fcir f:0 r:.01
  \move(.7 -1.4)\fcir f:0 r:.01
  \htext(3.5 .1){$n_1$}
  \htext(2.5 -.4){$n_2$}
  \htext(1.2 -1.9){$n_{N-1}$}
  \htext(.9 -2.4){$n_N$}
  \etexdraw
$$

\vspace{2.5cm}\noindent \textsc{\bf Classical densities and the
Groenewold operator} \vspace{.5cm}

In order to check our conclusions on the form of the mixed state
representing the SS, we use the embedding of classical functions
into quantum mechanic operators, develop by Groenewold
\cite{groenewold}. Basically, to define this quantization method,
it is used the fact that the Weyl map of equation (\ref{weyl}) has
a well defined inverse map. Then, given a classical density
$\rho_c$ the associated operator density is defined as follows
\bea
\hat\rho=\int{dpdq
\,\rho_c}\,\hat\Delta(q,p)\quad\hbox{where}\quad \hat\Delta(q,p)=
e^{i(q\hat p-p\hat q)}\,\hat P\, e^{-i(q\hat p-p\hat q)}
\eea
and $\hat P$ is the parity operator.

In this formalism, the calculation of expectation values corresponding to classical
average, takes the familiar quantum mechanic form $Tr(\hat A \hat \rho)$.
Also, it is easy to prove that for the Hilbert space basis $\{\ket{n}\}$
corresponding to eigenfunctions of the Hamiltonian of the single particle,
$\hat \rho$ is diagonal i.e. $\bra{m}\hat \rho\ket{n}= \delta_{(m,n)}\lambda_n$.

Consider in particular, the form of the SS density in the LLM framework,
\bea
\rho_c=\left\{
    \begin{array}{cc}
    {N\over(\pi R_{Q}^2)} & \quad r^2<R_Q^2 \\
    0 &    \quad r^2>R_Q^2
    \end{array}\right.
\eea
where by definition, integrating over the phase space gives
the total number of particles $N$. To make a clear connection with
previous results we recall that the Hamiltonian of a Harmonic
oscillator can be written as $H=w r^2/2$, where $r^2=\hbar\tau$
and we also define $\tau_Q=\hbar R_Q^2$ to obtain
\bea
\rho_c=\left\{
    \begin{array}{cc}
             {N\over\pi\hbar \tau_Q} & \;\;\tau<\tau_Q \\
             0  &  \quad\tau > 1
    \end{array}\right.
\eea
Then, it is not difficult to compute the
eigenvalues of $\hat \rho$ when expressed in terms of energy eigenvectors
$\ket{n}$, obtaining
\bea
\label{lam}
\bra{m}\hat \rho\ket{n} = \left\{
    \begin{array}{cc}
    {2N(-1)^m\over\tau_Q}\int_0 ^{\tau_Q} d\tau \,[e^{-\tau}L_{(n)}(2\tau)] & \quad m=n \\
             \\0  &  \quad m\neq 0
    \end{array}\right.
\eea
That should be plug in the final form of $\hat \rho$,
\bea
\hat \rho=\sum_{n=0}^\infty \lambda_n \ket{n}\bra{n}
\eea

\DOUBLEFIGURE{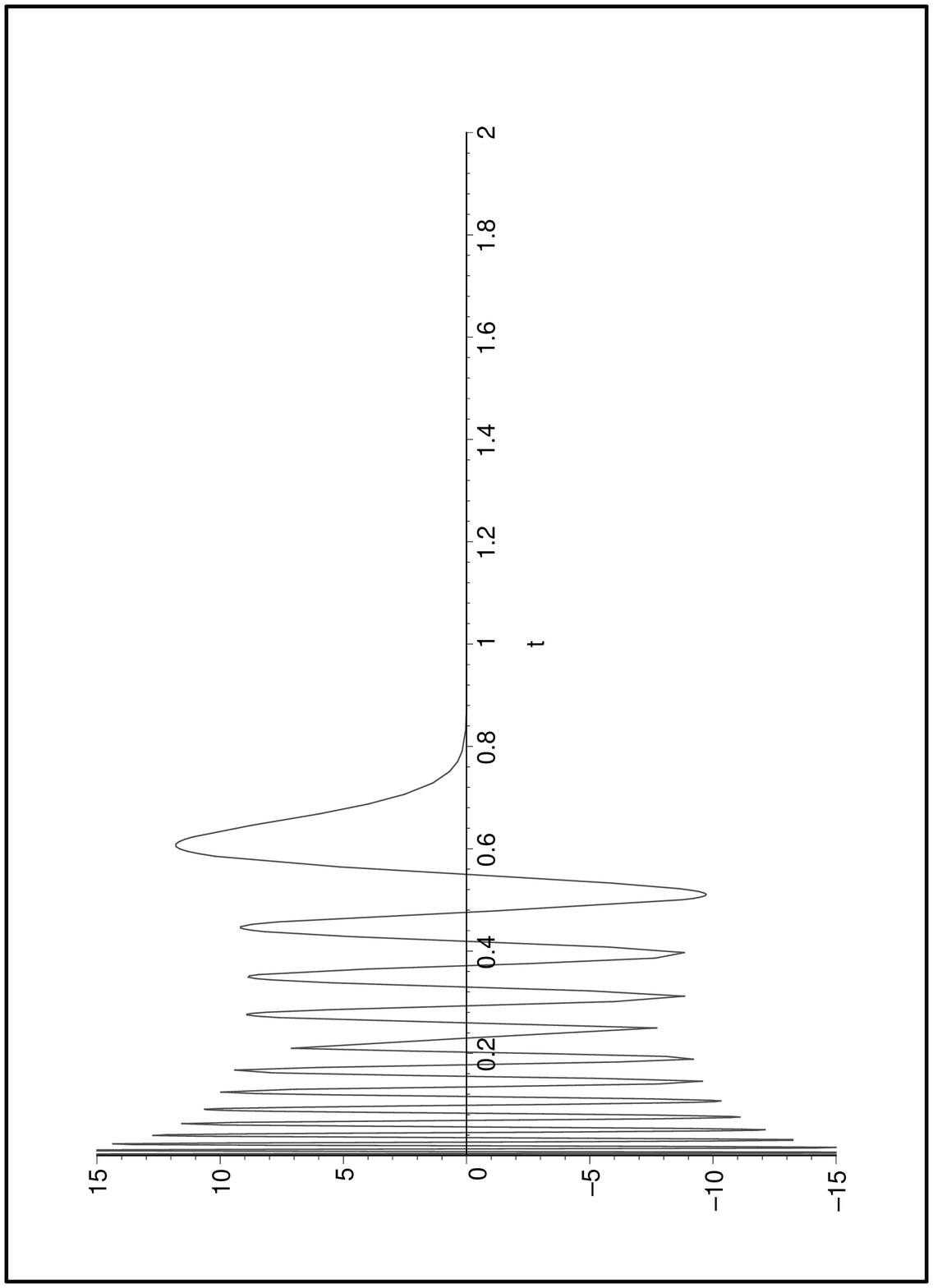,angle=-90,scale=.3}{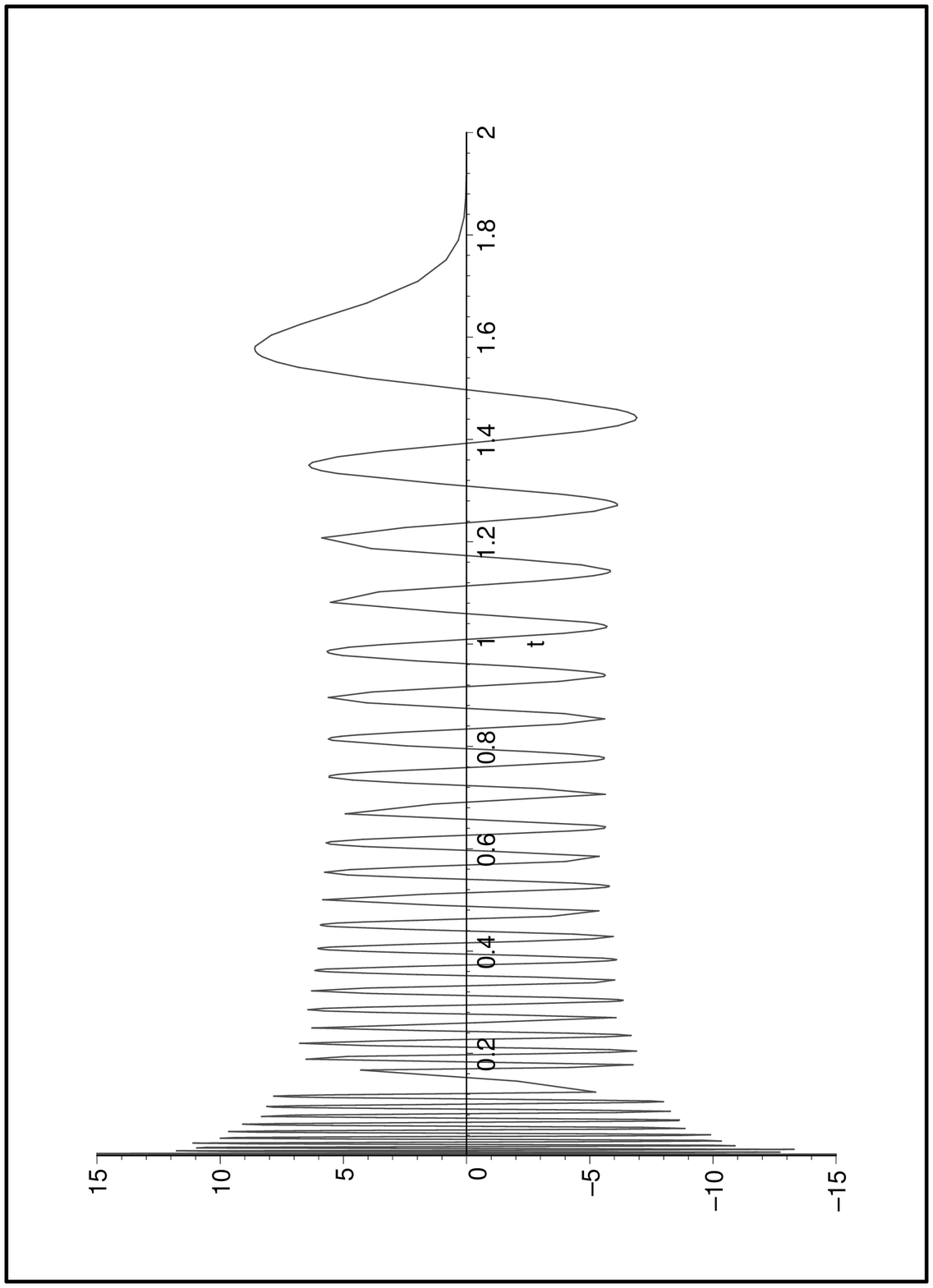,angle=-90,scale=.3}{The delta
distribution is center at $\tau=26/40$ and hence since in this normalization we
integrate up to 1, the integral is not zero}{The delta
distribution is center at $\tau=66/40$, and the integration gives zero}

At this point, it is important to note that $\lambda_n$ has a very
interesting behavior. For large N, $\lambda_n$
approaches zero if $n>N(1+Q)$! and approaches $1/(1+Q)$
if $n<N(1+Q)$. The reason behind this behavior is that the
integrand of equation
(\ref{lam}) tends to $1/(1+Q)$ times the delta function
with support in the interval $(0,1)$ if $n<N(1+Q)$ and
out of the interval if $n>N(1+Q)$ (see fig. 5 and fig. 6).

Summarizing, We have embedded the classical density for the SS
into a single particle quantum mechanic system. We have found that
the corresponding density operator is given by a mixed state that
is built using the hole tower of states $\ket{n}$. Nevertheless,
out of the this infinite set of $\lambda_n$, only those that have
$n<N(1+Q)$, give a sizable contribution. Hence we can approximate
our density operator as follows, \bea \hat
\rho\simeq\sum_{n=0}^{N(1+Q)} {1\over (1+Q)} \ket{n}\bra{n} \eea

Therefore, using our result of the previous sections that
the density operator of a single particle can be understood as the
density operator of N-fermions, we have found the density operator
of the SS, has as expected the form of a statistical
ensemble, that does not correspond to the equiprobable ansatz.
Note that, all the energy levels are populated with the same probability
in the single particle picture!, and that doest not corresponds to
an equiprobable ensemble in the N-fermion picture, but to a triangular
young diagram!.

\section{Discussion}

In this work adopt the perspective that GR is a mean-field
approximation of the microscopic structure defined by the true
ultraviolet degrees of freedom. In this approach, GR observables
are thermodynamical in nature, an should be understood in terms of
statistical mechanics of the ultraviolet degrees of freedom.

As an example of the above ideas, we focus on the simplest 1/2 BPS
sector of the $AdS_5/CFT$ duality. Here due to the simplicity of
the CFT sector, we are able to study the corresponding statistical
mechanics to give a rational foundation to the GR dual
observables.

We found necessary to use Weyl and Wigner formalism to incorporate
the phase-space picture into the quantum mechanics formalism. In
particular we found that the Wigner density correspondent to a
$N$-fermions pure state, can be understood as a mixed state of a
single particle. Then, all types of Wigner densities corresponding
to pure or mixed states can be recast in terms of mixed states of
the single particle.

Classical limit is surprisingly easy to take, the unpleasant
features of the Wigner density cancel out,
showing a clear picture where the classical point particles emerge as delta
functions. In particular we recover in this
form, the vacuum configuration, giant gravitons and in general al
the previous known configurations single out in GR that where built out of
concentric rings.

As a bonus of the above construction, we explain why the GR
description is linear in the droplets. Basically, each drop is
related to classical limit of mixed state, and the Wigner density
of the system is the sum or linear superposition of all mixed
states.

At last, we have analyzed the SS solution of GR from the
statistical mechanics point of view. We found that the
corresponding mixed state, is not describe by an equiprobable
distribution of states with fixed energy, but rather a
distribution where only those states with almost triangular Young
diagram are present. Nevertheless, when the distribution is
rewritten in terms of the single particle states, we obtain an
equiprobable distributions as final result. It seems to us, that
in this sector the theory is telling that the natural variables to
be used are the single particle ones.

The sector study in this work, has received some attention lately,
in particular research on its quantum structure and consequences
to the GR picture result in a chronology protection mechanism
based on the Pauli exclusion principle \cite{Caldarelli:2004mz}.
Also, the direct quantization of the LLM solutions has been
studied. Here, the collective coordinate quantization can be
reproduced by the D3-branes effective action in the LLM background
\cite{Mandal:2005wv}. A more canonical scenario was considered in
\cite{r}, where a pure GR approach with no mention to stringy
tools was undertaken. In this case, the bosonization of the
underlying microscopic femionic theory is recovered. Nevertheless,
from our point of view, it is hard to see how the full quantum
structure would be recovered once the coarse graining is done in
this case not to mention in other more general cases. At last
"Bubbling AdS" have been exported to other systems, like the D1/D5
case \cite{l}, and again has brought more understanding on the
quantum structure of GR \cite{Boni}. Unfortunately, in this case
we do not have a good idea of what is the equivalent of the
$N$-fermions picture for the microscopic degrees of freedom.
Definitively more study should be done in this directions.

 We believe that this work is "a nice example" where the
 underlying structure of space-time is exposed, and that
 with some luck, nature works in a similar
fashion. In any case, there are plenty of things to uncover in
this context, not to talk about the general case.


\section*{Acknowledgments}
\small We thank  M. Panareda for useful discussions and enlighting
converzations. This work was partially supported by INFN, MURST
and by the European Commission RTN program HPRN-CT-2000-00131, in
which M.~B. and P.~J.~S. are associated to the University of
Milan. \normalsize

\end{document}